\documentclass[fleqn,usenatbib]{mnras}

\usepackage{newtxtext,newtxmath}
\usepackage[utf8]{inputenc}
\usepackage[T1]{fontenc}
\usepackage{graphicx}	
\usepackage{amsmath}	
\usepackage{natbib}
\usepackage{rotating}
\usepackage{caption}
\usepackage{hyperref}
\usepackage{url}
\usepackage{longtable}
\usepackage{tablefootnote}
\usepackage{lipsum}    
\usepackage{booktabs}
\usepackage{siunitx}
\usepackage{fix-cm}
\usepackage[table]{xcolor} 
\newcommand{\msun}{\mbox{$\,{\rm M}_\odot$}}
\newcommand{\kms}{\,\mathrm{km\,s^{-1}}} 
\newcommand{\kpc}{\,\mathrm{kpc}}     

\title[The ONC as the precursor of the Pleiades and the Hyades clusters] {Are the ONC, Pleiades, and Hyades snapshots of the same embedded cluster?}

\author[Safaei et al.]
{Ghasem Safaei$^{1}$\thanks{
		E-mail: \mbox{gh.safaei@iasbs.ac.ir} },
	Hosein Haghi$^{1, 2, 3}$\thanks{
		E-mail: \mbox{haghi@iasbs.ac.ir} },
	Akram Hasani Zonoozi$^{1, 3}$ and Pavel Kroupa$^{3, 4}$ 
	\\
	$^{1}$Department of Physics, Institute for Advanced Studies in Basic Sciences (IASBS), 444 Prof. Sobouti Blvd., Zanjan 45137-66731, Iran\\
    $^{2}$School of Astronomy, Institute for Research in Fundamental Sciences (IPM), PO Box 19395 - 5531, Tehran, Iran\\
	$^{3}$Helmholtz-Institut f\"ur Strahlen-und Kernphysik (HISKP), Universit\"at Bonn, Nussallee 14-16, D-53115 Bonn, Germany\\
	$^{4}$Charles University in Prague, Faculty of Mathematics and Physics, Astronomical Institute, V Hole\v{s}ovi\v{c}k\'ach 2, CZ-180 00 	Praha 8, Czech Republic\\
}

\date{Accepted XXX. Received YYY; in original form ZZZ}

\pubyear{2025}

\begin{document}
\label{firstpage}
\pagerange{\pageref{firstpage}--\pageref{lastpage}}
\maketitle

\begin{abstract}
Using direct $N$-body simulations, we investigate the initial conditions and evolution of a star-forming region resembling the Orion Nebula Cluster (ONC) with the advanced \textsc{NBODY6} code. By varying the initial conditions, we aim to identify a model that closely aligns with observed parameters such as the half-mass radius, core radius, and total mass. Additionally, we examine the cluster's evolution over 800 Myr to determine whether it could reproduce the present-day properties of the Pleiades and Hyades along its evolutionary path.  Under the influence of a Milky Way-like tidal field, the ONC experiences significant mass loss, primarily due to rapid gas expulsion, retaining approximately 47\% of its initial 4200 stars by about 100 Myr and only 9\% by about 700 Myr. These evolutionary stages closely match the properties of the Pleiades and Hyades, suggesting that an ONC-like cluster may have been their precursor. Additional models with varying degrees of primordial mass segregation indicate that the ONC likely had an initial half-mass radius of 0.2–0.3 pc, a total mass of 1200 - 2000 M$_\odot$, and a high degree of mass segregation. Models with an initial stellar count of about $N_{\text{in}} \approx  4 \times 10^3 - 5 \times 10^3$,  rich in binaries and exhibiting mass segregation, show excellent agreement with observed cluster properties.

\end{abstract}

\begin{keywords}
	methods: numerical - stars: initial mass function - globular clusters:
	individual: ONC, Pleiades, and Hyades.
\end{keywords}

\section{Introduction}

Most Galactic field stars are not formed in isolation; rather, they originate within star clusters (SCs) embedded in giant molecular clouds that are initially compact structures \citep{KroupaAH01, Lada03, Gieles12, Marks12}. The study of present-day characteristic parameters of very young embedded SCs that still form provides valuable insight into the initial conditions and the processes that govern the dynamical evolution of SCs post-formation.

The early evolution (first $\approx$ 10 Myr) of young, gas-embedded SCs is primarily dominated by the removal of residual gas from star formation, driven by feedback from massive stars through UV radiation, stellar winds from OB stars, and supernova explosions. Consequently, star clusters become supervirial in a timescale shorter than or comparable to the crossing time of stars within the cluster. The subsequent dynamical changes in the cluster are significantly affected by this gas loss. The impact of early gas expulsion (GE) on the dynamical evolution of star clusters has been investigated by \cite{Baumgardt07} and \cite{Haghi15}, who used a large set of N-body integrations. Their findings revealed that the initial conditions for GE play a crucial role in determining the astrophysical characteristics of the star cluster, especially during its formative Myr. The fraction of high-mass stars significantly influences the dynamical evolution of star clusters \citep{Wang2021}. \citet{Haghi20} showed that clusters with a top-heavy initial mass function (IMF) dissolve faster and experience greater half-mass radius expansion compared to those with a canonical IMF under identical initial conditions.
Furthermore, \cite{Zonoozi11, Zonoozi14} found that a shallower slope of the mass function observed in Pal 4 and Pal 14, specifically in the mass range of 0.55 - 0.85 $\msun$, can be replicated through the GE effect in initially mass-segregated clusters.

To understand the formation of star clusters, the origins of their stars, and the initial conditions at their birth, it is essential to study very young massive clusters (YMCs), such as the Orion Nebula Cluster (ONC), the Galactic cluster NGC 3603, and R136 in the Large Magellanic Cloud (LMC). Nearby open clusters like the Pleiades and the intermediate-aged Hyades also provide valuable insights.

The YMCs are typically very compact, with sizes of $\leq 1$ pc. In contrast, open star clusters, such as the Pleiades, are much more expanded and typically span several parsecs. After the gas expulsion phase, a cluster reforms due to energy equipartition during the embedded or expanding phase \citep{KroupaAH01}, ultimately filling its Roche lobe within the Galactic tidal field. This process results in an expanded core radius; for example, the ONC has a core radius of approximately 0.2 pc, while the Pleiades core radius is approximately 1.3 pc, and after the re-vitalization the cluster evolves through secular evaporation \citep{BM03}. Since both YMCs and open clusters must initially form in a very compact state from molecular cloud clumps, a key question arises: how do these clusters, which start so compact, expand to reach the much larger sizes observed in older clusters? This size difference provides an important constraint on the formation and evolutionary processes of star clusters.

Two primary scenarios have been proposed for the formation of star clusters: either they form hierarchically from smaller substructures, or they emerge through a monolithic collapse within a dense molecular cloud clump. Very young massive star clusters, such as NGC 3603 and R136, serve as ideal examples to test these formation models. \citet{Banerjee15} investigated the hierarchical assembly of sub-clusters as a possible mechanism for forming these very young massive clusters. Their findings indicate that the hierarchical merging scenario is unlikely, as there is insufficient time to build such clusters from smaller structures.  Instead, they showed that the typical present-day density ($\rho \approx 10^4 -10^5~\rm M_{\odot} pc^{-3}$), very young age ($\approx 1-3  \rm ~Myr$) and near-spherical core-halo morphology of gas-free massive clusters, like R136 and NGC 3603, dictate an episodic or monolithic formation of star clusters. This suggests clusters form either in situ or through a prompt assembly phase, followed by a phase of violent gas dispersal. This is supported by the analysis of the spatial distribution of molecular cloud clumps within which form embedded clusters and very young clusters \citep{Zhou2024, Zhou2025}. A very good understanding of the formation and evolution of very young clusters has been achieved \citep{MarksKDP12}.

The ONC is the nearest star-forming region in which ionising (massive, $m> 10 M_{\odot}$) stars have been forming, located approximately 500 pc from Earth. At an age of just 2.5 Myr \citep{Hillenbrand97}, it is one of the youngest known clusters, containing around 4000 stars distributed across a region with a diameter of 5 pc. The cluster exhibits a higher-than-expected velocity dispersion among its stars, which is unusually large given its known mass \citep{Jones88}. This suggests that the ONC is expanding, likely due to the rapid expulsion of residual gas left over from the star formation process. Despite its large velocity dispersion, which indicates that the ONC might be contributing to an unbound association, N-body simulations reveal that the inner cluster is on the verge of becoming a bound system. This occurs despite significant mass loss caused by its massive stars and the influence of the Galactic tidal field \citep{Kroupa00}.

\cite{KroupaAH01} studied the present-day structure of the ONC, and the Pleiades open clusters by $N$-body simulation for 100 Myr. These computations were the first to study the formation of open star clusters, using high-precision N-body modelling,  including all physical effects like explosive GE, Galactic tidal field, and mass loss from stellar evolution and a very high initial binary fraction. They used two initial models and derived the key properties of the clusters. For both models, they found good agreement with the observational data using Aarseth's $N$-body code that was modified for the GE effect. As the ONC is likely to be expanding rapidly and probably having lost its outer stars, \cite{KroupaAH01} showed that a relatively sparse cluster like the Pleiades must once have been in a state similar to the ONC. In other words, an ONC-like cluster may have been a precursor of the Pleiades as the ONC resembles the Pleiades cluster to a remarkable degree after 100 Myr of stellar and dynamical evolution. Here in this paper, we aim to investigate the likely fate of the ONC and assess whether the clusters Pleiades and Hyades can be in the evolutionary path of an ONC-like cluster using the state-of-the-art \textsc{NBODY6} code, incorporating the gas in an embedded cluster as a time-evolving background potential. Calculations are performed for clusters initially set up with very compact configurations and then evolved to compare with the observed properties of the ONC, Pleiades, and Hyades clusters, respectively, over time.

The paper is organized as follows. In Section \ref{Sec: Observations} the observational data on the ONC, Pleiades, and Hyades, including the total mass, the half-mass radius, and the tidal radius, are presented and will later be compared to our simulations using the above constraints. In Section \ref{Sec:Description of the models} we describe the set-up of the initial $N$-body models. This is followed by a presentation of the main simulation results in Section \ref{Sec:Results}. Finally, the results and conclusions are discussed in Section \ref{Sec:Conclusions}.

\section{Observational data}\label{Sec: Observations}

We compare the results of our numerical modeling of the ONC, Pleiades, and Hyades clusters that are also visible to the naked eye at a similar location in the night sky with observational spectroscopic and photometric data from the literature. In this section, we summarize the available observational constraints for these clusters, as outlined in Table \ref{tab: observational-results-OPH-table1}.

\begin{figure}
\includegraphics[width=9 cm]{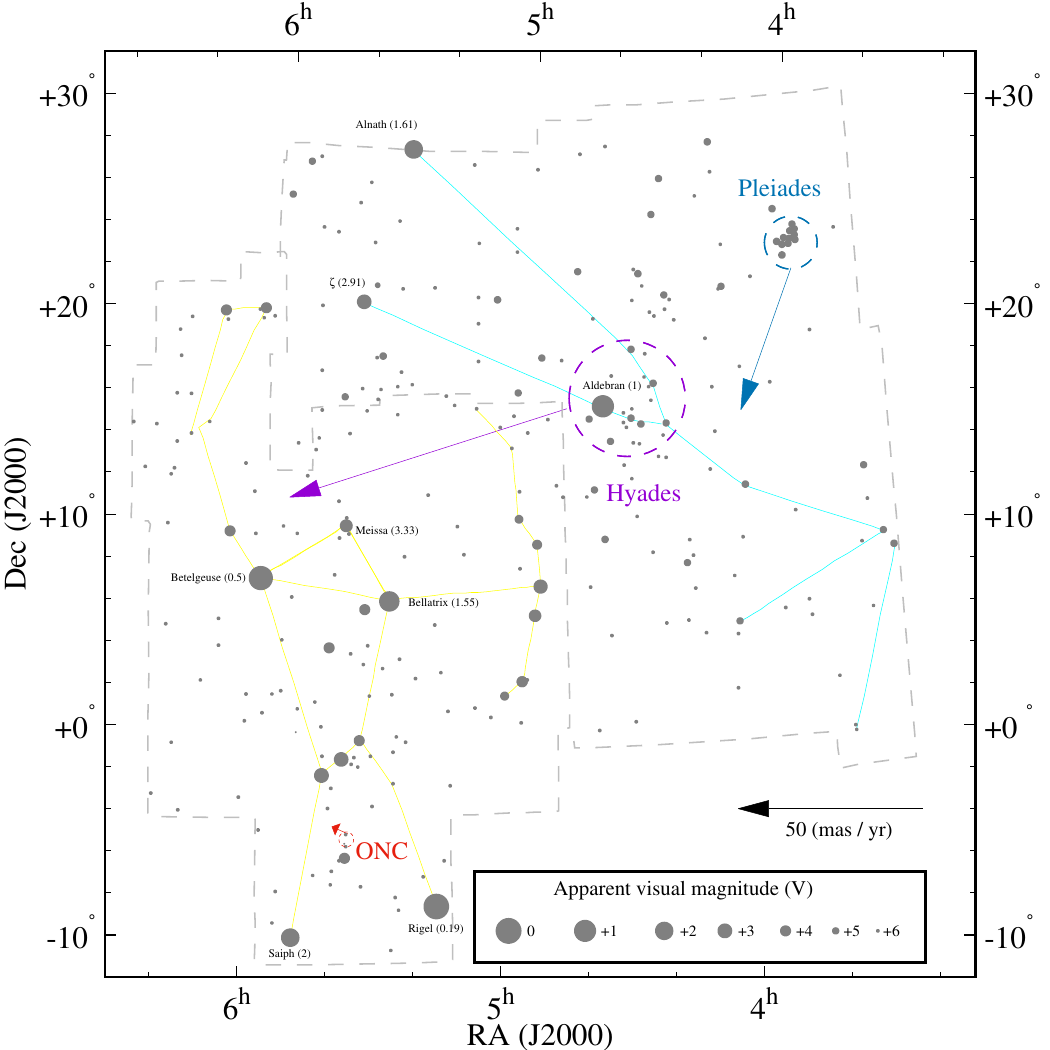}
\caption{The constellation of the Orion and the Taurus with three open clusters: the ONC, the Pleiades, and the Hyades. See text for further details.
}
\label{fig:constellation-onc-tau}
\end{figure}

Figure \ref{fig:constellation-onc-tau} displays the Orion and Taurus constellations, with red, blue, and purple dashed circles highlighting the regions associated with the ONC, Pleiades, and Hyades, respectively. The horizontal axis represents right ascension (J2000), while the vertical axis represents declination (J2000). Filled black circles denote stars visible to the naked eye, with their sizes proportional to their apparent visual magnitudes, following a linear relationship between point size and magnitude. Yellow and cyan lines trace the constellations' patterns, and grey dashed lines divide the fields. Arrows represent the mean proper motion vectors for each cluster, indicated by the average $\mu^{*}_{\rm RA}$ and $\mu_{\rm Dec}$. A black arrow in the bottom-right corner indicates a proper motion scale of 50 mas/yr. \citep{vanleeuwen09, Galli17, Dzib21, Ebrahimi22}.

\subsection{ONC}\label{subsec:onc}

The ONC, is the closest star-forming region in which ionising stars of spectral type O have formed located at about $ 403 ^{+7}_{-6}$ pc away from us \citep{Kuhn19}. It is one of the youngest known clusters, containing approximately 2700 stars spanning the full mass spectrum, from very low-mass brown dwarfs to massive O-type stars \citep{Kroupa18, Wang19, Jerabkova19}. These stars are distributed across a region with a diameter of about 5 pc.

The ONC is known to be a very young ($<$ 0.1 - 2 Myr, with a mean age of $<$ 1 Myr) star-forming region of significant interest \citep{Hillenbrand97, Hillenbrand98}. However, \cite{Beccari17} presented the results of an optical photometric survey of the ONC in a region of $12^{\circ}\times 8^{\circ}$ area using OmegaCAM and the 2.6 m VLT Survey Telescope (VST) at Cerro Paranal in Chile. They discovered three well-separated sequences of pre-main-sequence stars in the color-magnitude diagram. Using Gaia DR2 in combination with the photometric OmegaCAM catalogue and utilizing the Pisa stellar isochrones,  \cite{Jerabkova19}  determined the absolute magnitudes of the ONC members and confirmed the existence of these three putative stellar populations with the best fitting parameters: 4.5$_{-1.2}^{+1.5}$, 2.1$_{-0.4}^{+0.5}$, and 1.4$_{-0.2}^{+0.3}$ Myr implying three separated episodes of star formation. We took the mean age of $\approx 2.1 \pm 1$ Myr for the ONC in this paper. The three populations are probably the result of gas-inflowed driven star formation modulated by the repeated ejection of ionizing stars \citep{Kroupa18}. 

The inner ONC which is known as the Trapezium region has the stellar number density $\approx 10^4$ {pc}$^{-3}$ and is the densest nearby star-formation region \citep{Hillenbrand98, McCaughrean94}.  \cite{Hillenbrand97}  used optical and near-infrared stellar source counts and derived the total observed stellar mass and the number of stars of the ONC within (r $<$ 2.06 pc) to be $M_{star}\approx$ 1850 M$_\odot$ and N$_{star}$ = 2260, respectively.  
\cite{Dario14} analysed the ONC, utilizing the latest stellar membership over a large spectrum of wavelengths (optical, X-ray, and JHK) based on a collection of stellar catalogues. They summarized the number of all individual sources and total stellar mass of the ONC within 2 pc using the best-fit of density profile parameters to the results of catalogues (optical photometry, near-infrared photometry complemented with 2MASS, Spitzer, and X-ray) and  found N$_{star}$~=~2745 and M$_{star}$ = ~1539~M$_\odot$.  \citet{Pavlik19} reanalyzed these data in view of the ONC having been fully mass segregated.

To determine the radius from the centre of the cluster that contains half of the total mass of $\approx$ 2260 observed stars, \cite{Hillenbrand98} assumed the limiting radius of the ONC of about 2.06 pc and reported a half-mass radius ($r_{hm}$) of 0.8 pc. They defined the core radius ($r_0$) as the distance at which the surface density of the cluster drops to approximately half of its central value, estimating it to be between 0.16 and 0.21 pc.

\subsection{Pleiades}\label{subsec:Pleiades}
The Pleiades is an open cluster located in the constellation of Taurus at a distance of about 136 pc. Assuming approximately $110.0 ^{+1.4}_{-0.9}$ Myr old containing around 1059 members  \citep{Gaia2018}. \cite{Pinfield98} studied 1194 candidate members of the Pleiades cluster located within a $3^{\circ}$ survey radius area in approximately virialized conditions and using the King fit profile method (with four different mass bins centered at 5.2, 1.65, 0.83 and 0.3 $\msun$) they found the stellar total mass to be $\approx$ 740 M$_\odot$. They also calculated the half-mass radius to be $\approx$ 3.66 pc.

\subsection{Hyades}\label{subsec:Hyades}
The Hyades is the closest open cluster to the Sun, located about 46 pc away, making it a popular subject of study among astronomers. As one of the best-studied star clusters, it resides in the Taurus constellation and contains around 400 potential members, with a total stellar mass of roughly 300 - 400 M$_\odot$, spanning an area of 20 degrees in the sky \citep{Perryman98}. Based on the Tycho-Gaia Astrometric Solution (TGAS) for 103 star members, \cite{Gaia2017} reported a distance of $\approx$ $46.75 \pm 0.46$ pc. A more recent study using Gaia Data Release 2 identified 480 stars as members of the Hyades cluster. By analyzing the Hertzsprung-Russell diagram, the age of the cluster was determined to be $\approx$ 794.0 Myr with a metallicity of Z = 0.02 dex, at a distance of $\approx$ 47.62 pc \citep{Gaia2018}.\\

\begin{center}
\begin{table*}
\centering
\begin{minipage}{\textwidth}
\caption{ Observed and derived properties of the ONC, Pleiades, and Hyades, which are used for comparison with our simulations. The first column lists the cluster age (in Myr), while columns 2 to 8 present the following properties: the three-dimensional tidal radius, the number of stars, the total mass, the half-mass radius, the core radius, the distance from Earth, and the corresponding reference.}
\begin{tabular}{cccccccc}
\hline
 \textbf{Age} & $\mathbf{r_t}$ & \textbf{N} & \textbf{Mass} & $\mathbf{r_h}$ & $\mathbf{r_c}$ & \textbf{Distance} & \textbf{Reference} \\ 
\text{[Myr]} & [pc] & & $\text{[$\msun$]}$ & [pc] & [pc] & [pc] & \\ 
$(1)$ & $(2)$ & $(3)$& $(4)$ & $(5)$ & $(6)$ & $(7)$ & $(8)$ \\
\toprule
\multicolumn{8}{c}{\textbf{ONC}} \\ 
\midrule
 $2.6 \pm 1$ & - & - & - & - & - & - & \protect\cite{Jerabkova19} \\ 
 - & - & - & - & - & - & $403 \pm 6$ & \protect\cite{Kuhn19} \\ 
 - & - & 2399\textsuperscript{\textdagger} & - & - & - & - & \protect\cite{Pavlik19} \\ 
 - & - & 1648\textsuperscript{\textdagger}& - & - & - & - & \protect\cite{Schlafly19} \\ 
  $2 \pm 1$ & & $2745 \pm 300$ & - & - & - & - & \protect\cite{DaRio17} \\ 
 $2.0 \pm 0.7$ & - & - & - & - & - & - & \protect\cite{Beccari17} \\ 
 - & - & - & - & - & - & $388 \pm 5$ & \protect\cite{Kounkel17} \\ 
 - & $18 \pm 1$\textsuperscript{\textasteriskcentered} & $2745 \pm 300$ & $1539 \pm 100$ & $0.9 \pm 0.1$ & - & - & \protect\cite{Dario14} \\ 
 - & - & 1743\textsuperscript{\textdagger} & - & - & - & - & \protect\cite{Dario12} \\ 
 - & - & 2621\textsuperscript{\textdagger} & - & - & - & - & \protect\cite{Dario09} \\ 
 - & - & - & - & - & - & $414 \pm 7$ & \protect\cite{Menten07} \\ 
 - & $17 \pm 3$\textsuperscript{\textasteriskcentered} & $ 1740 \pm 170$ & $ 1100 \pm 200$ & - & - & - & \protect\cite{Lada03} \\ 
 - & 19\textsuperscript{\textasteriskcentered} & 2260 & 1800 & 0.8 & $0.16-0.21$ & - & \protect\cite{Hillenbrand98} \\ 
 $1.1 \pm 0.8$ & - & - & - & - & - & - & \protect\cite{Hillenbrand97} 
 \\ 
\hline
\multicolumn{8}{c}{\textbf{Pleiades}} \\ 
\hline
 - & $13.69 \pm 0.3$ & 1639 $\pm$ 90 & 947 $\pm$ 87 & - & - & - & \protect\cite{Hunt24} \\
 - & $13.54 \pm 0.27$ & 1448 $\pm$ 88 & 807 $\pm$ 48 & 2.96 $\pm$ 0.09 & - & - & \protect\cite{Ebrahimi22} \\
 - & 11.8 & $1177 \pm 9$ & 850 & 4.4\textsuperscript{\S} & - & - & \protect\cite{Meingast21} \\ 
  - & 11.5 & $1387 \pm 10$ & 770 & - & - & - & \protect\cite{Roser20} \\ 
 $110 \pm 1$ & - & $1059 \pm 5$ & - & - & - & $135 \pm 5$ & \protect\cite{Gaia2018} \\ 
  - & $18-19$ & 2272 & $1408-3571$ & - & $1.3-3.0$ & - & \protect\cite{Olivares18} \\
 $130 \pm 5$ & - & - & - & - & - & - & \protect\cite{Gaia2017} \\ 
 - & - & - & - & - & - & $134 \pm 6$ & \protect\cite{Gaia2016} \\  
 - & - & 2109 & - & - & - & - & \protect\cite{Bouy15} \\ 
- & - & - & - & - & - &  $136 \pm 1$ & \protect\cite{Melis14} \\ 
 - & 19.5 & $1244 \pm 35$ & 939 & - & $2.2 \pm 0.4$ & - & \protect\cite{Converse10} \\ 
$120 \pm 2$ & - & - & - & - & - & $120 \pm 1$ & \protect\cite{vanleeuwen09} \\ 
 $120 \pm 1$ & $10.9 \pm 0.9$ & - & $468 \pm 12$ & - & - & $130 \pm 8$ & \protect\cite{Piskunov08} \\
 - & - & - & - & - & - &  $133 \pm 1$& \protect\cite{Soder05} \\
 - & 16.0 & - & 800 & - & $2.35-3.0$ & - & \protect\cite{Adams01} \\ 
 - & 17.5 & 1400 & $530-2900$ & 1.9\textsuperscript{\S} & 1.4 & - & \protect\cite{Raboud98} \\ 
 - & 13.1 & 1195 & 735 & 3.66 & $0.91-2.91$ & - & \protect\cite{Pinfield98} \\ 
\hline
\multicolumn{8}{c}{\textbf{Hyades}} \\ 
\hline
 - & $8.06 \pm 0.2$ & 569 $\pm$ 90 & 193 $\pm$ 42 & - & - & - & \protect\cite{Hunt24} \\
 - & 12.51 $\pm$ 1.09 & 582 $\pm$ 110 &649 $\pm$ 170& 3.97 $\pm$ 0.15 & - &-& \protect\cite{Ebrahimi22} \\ 
 $640^{+67}_{-49}$ & $9.0 \pm 0.9$ & 385 $\pm$ 38 & $275 \pm 15$ & 4.1 & 3.1 & $47.03 \pm 0.2$ & \protect\cite{Lodieu19} \\ 
$794 \pm 1$& - & 480 & - & - & - & $48 \pm 3$ & \protect\cite{Gaia2018} \\ 
$794 \pm 2$ & - & - & - & - & - & $47 \pm 4$ & \protect\cite{Gaia2017} \\ 
 - & 9.0 & 364 & 435 & 4.1 & 3.1 & - & \protect\cite{Roser11} \\ 
 - & - & - & - & - & - & $46 \pm 5$ & \protect\cite{vanleeuwen09} \\ 
$625 \pm 50$ & 10.0 & 282 & 400 & 5.7\textsuperscript{\S} & 2.9 & $46 \pm 6$ & \protect\cite{Perryman98} \\ 
 - & 10.5 & 393 & $410-480$ & - & - & - & \protect\cite{Reid92} \\ 
 - & 9.5 & 400 & 300 & 3.6 & - & - & \protect\cite{Pels75} \\ 
\bottomrule
\end{tabular}
\label{tab: observational-results-OPH-table1}

\vspace{0.1cm}
\textsuperscript{\textdagger} The study provides candidate star counts derived from uploaded tables, accessible via the  \href{http://cdsportal.u-strasbg.fr}{CDS}  portal for each reference. \\
\textsuperscript{\textasteriskcentered} The tidal radii of the ONC were calculated using Equation (\ref{equ:rt}) based on the reported mass values.\\
\textsuperscript{\S} Their $r_h$ calculation method differs from ours in N-body simulations.    Note that differences in $r_h$ values arise due to varying calculation methods used in the literature, particularly the inclusion of tidal tails or different stellar mass ranges, which are not directly comparable to our N-body results.
\end{minipage}
\end{table*}
\end{center}
%
\section{Description of \texorpdfstring{$N$-body}{N-body} models and initial conditions}\label{Sec:Description of the models}
In this section, we discuss the properties of the code and also describe our choices for the initial total mass, half-mass radius, the initial stellar mass function (IMF), and the orbital parameters of the models that form the initial compact embedded SCs. Finally, in Table \ref{tab:initial-conditions-ABC-table2} we summarize the initial cluster parameters used for the modelling of the ONC.\\

For the numerical simulations, we utilize the state-of-the-art \textsc{Nbody6} code \citep{Aarseth03, Nitadori12}, which has been modified for use with Graphics Processing Units (GPUs). This code is specifically designed to accurately integrate many-body systems and compute the dynamical evolution of star clusters (SCs) over their lifetimes. \textsc{Nbody6} employs fourth-order Hermite integration to calculate the orbit of each cluster member as a function of time.

For each model, we adopt the three-part power-law canonical initial mass function (IMF) to randomly generate the initial distribution of stellar masses between $ m_{min}$ = 0.01 $ M_\odot $ and $ m_{max} $ = 50 $ M_{\odot} $ in the form of \citep{Kroupa01}.

\begin{eqnarray}
\,\,\,\,\frac{dN}{dm}\propto m^{-\alpha},\,\,\,\, 
\,\left\{
\begin{array}{ll}
\alpha _1 = +0.3, \hspace{0.25 cm}\hspace{0.25 cm} 0.01 \leq \frac{m}{M_{\odot}}<0.08,\\
\alpha _2 = +1.3, \hspace{0.25 cm} \hspace{0.25 cm} 0.08 \leq \frac{m}{M_{\odot}}<0.50,\\
\alpha _3 = +2.3, \hspace{0.25 cm} \hspace{0.25 cm} 0.50 \leq \frac{m}{M_{\odot}}<50.0.
\end{array}
        \right. \label{MF}
\end{eqnarray}

In all models, we assumed initial binary populations. The initial binary fraction, $B_F$, is defined as
\begin{eqnarray}
B_{F} = \frac{N_{bin}}{(N_{bin}+N_{sing})},\label{equ:bf}
\end{eqnarray}
where $N_{bin}$ and $N_{sing}$ represent the number of binary and single-star systems, respectively. For all models,  $B_F \approx 1$ (see Table \ref{tab:initial-conditions-ABC-table2}). For binary component pairing, the period distribution for stars of mass $m > 5 \msun$ was adopted from \cite{Sana12} and \cite{Oh16}. The mass ratio distribution ($0.1<q=\frac{m_2}{m_1}<1.0$) for primary masses  $m>5 \msun$, followed a uniform distribution based on \cite{Sana12} and \cite{Oh15}, while random pairing was applied for other stars using the same references ($m_2$ is the mass of the secondary). The semi-major axis distribution for all models was derived from the period distribution provided by \cite{Kroupa95a, Kroupa95b}. Additionally, the latest updates to the initial binary population distribution of SCs, as presented by \cite{Belloni17}, was implemented in the updated version of MCLUSTER \citep{Wang19}. See also \citet{Kroupa2025}  for a review. 

To model the evolution of single and binary stars from the zero-age main sequence to their late phases (remnants) along the Hertzsprung-Russell diagram (HRD), the SSE/BSE routines are incorporated into the \textsc{Nbody6}  code. These routines are based on analytical fitting functions and support a broad range of metallicities (from 0.0001 to 0.03) developed by \cite{Hurley00} and \cite{Hurley02} to produce a realistic CMD.  In these simulations, we adopted the standard solar metallicity value (Z = 0.02 dex or [Fe$/$H] = -1.41) for all our models.

To account for the high velocities of ejected remnants from massive stars, the \textsc{Nbody6} code assigns kick velocities to these remnants using a Maxwell velocity distribution. However, this aspect is not the focus of the current study. Given the evidence suggesting that heavy remnants escape from the ONC \citep{Hillenbrand97, Kraus09, Subr12}, we adopt an initial Maxwellian velocity distribution for black holes and neutron stars with $\sigma_{\rm BH/NS} = 190$ km/s, corresponding to a zero retention fraction.  White dwarfs (WDs) represent the final evolutionary stage of low-mass stars. During their post-main sequence evolution, WDs experience non-spherical mass loss, which results in an isotropic recoil velocity for the WD remnant. The strength of the "natal kick" a WD receives can influence the amount of dark remnants in clusters. For all types of white dwarfs (Helium, Carbon-Oxygen, Oxygen-Neon), we use  $\sigma_{\rm WD} = 5$  km/s as inferred from observations and consistent with values used in previous studies \citep{Fellhauer2003,Davis2008, Fregeau2009}. To configure the initial conditions for the various models, we employed the publicly available tool MCLUSTER \citep{Kupper11}.

The GE effect is considered to be an external background potential and a reduction in the cluster potential perturbs the equation of motion of stars. The spatial distribution of the gas is similar to that of stars with the same initial half-mass radius. Also, the mass of the gas decreases exponentially at times $t \geq 0.6 $ Myr (for all models, the delay time is considered to be $ 0.6~ \rm Myr$ which corresponds to the effective lifetime of the ultra-compact HII region) as in the following equation,
\begin{eqnarray}
 M_{\rm gas}(t)=M_{\rm gas,0} e^{-(t-0.6)/t_{\rm d}}, \label{equ:gas-mass}
\end{eqnarray}
where the time-scale is chosen to be $t_{\rm d}[Myr]=0.1~r_{\rm hi} [pc]$ \citep{KroupaAH01,Brinkmann17} and $M_{\rm gas,0}$ is the initial gas mass.

The star formation efficiency (SFE), defined as $SFE = \frac{M_{\star}} { M_{\star}+M_{\rm gas} }$, is estimated from observational results to be approximately $SFE \approx 0.33$ \citep{Lada03, Megeath16}.
\cite{Boily03}, using an analytical approach, and \cite{Brinkmann17}, based on N-body simulations of star clusters, both concluded that a higher initial SFE increases the number of stars that remain bound to the cluster. In simulations conducted by \cite{Baumgardt07}, initial SFE values ranging from 10\% to 75\% were considered, and it was found that clusters formed with SFE $\geq$ 33\% could survive gas expulsion (GE) under instantaneous gas removal conditions. Additionally, \cite{Dario14} analyzed the ONC across multiple wavelengths using a collection of catalogs to derive its mass density profile. Their findings suggested that models with an initial density profile resembling the present-day ONC distribution likely formed with an efficiency between 30\% and 50\%.

Initially, the stellar positions and velocities align with a Plummer density profile \citep{Plummer11} up to a cut-off radius of twice the cluster’s tidal radius, $r_t$. The star velocities are adjusted to ensure the cluster maintains virial equilibrium with the external tidal field.    One of our models begins with the King profile, setting the initial King's parameter at $W_0$ = 7. To initialize embedded star clusters, we adopted the mass-radius relation from \cite{Marks12}, expressed as:
\begin{eqnarray}
 \frac{r_{\rm hi}}{ pc}=0.1 \times \left(\frac{M_{\rm ecl}}{\msun}\right)^{0.13},
\label{equ:marks-kroupa-relation}
\end{eqnarray}
where $M_{\rm ecl}$ is the initial mass of the cluster and $r_{\rm hi}$ is the initial half-mass radius of the star cluster. For the three models indicated in Table \ref{tab:initial-conditions-ABC-table2} (M3.6k-S1, M3.7k-S1, and M3.7k-S1-K), the initial half-mass radius is taken from \cite{KroupaAH01}.

Recent investigations indicate that the ONC was formed in a completely mass-segregated way \citep{Pavlik19}. We also consider primordial mass segregation (PMS) using the MCLUSTER code for the initial setup of our simulations. As a result of energy equipartition in PMS models, the most massive stars are more centrally concentrated to the lowest level orbits than the lower mass stars. By choosing the degree of segregation between 1 and 0, where 1 means completely segregated and 0 means unsegregated, we can determine the degree of PMS. 
The present study examines the role of PMS in shaping the structural properties of star clusters. Rather than focusing on present-day mass segregation, we systematically varied initial PMS within acceptable ranges. This allowed us to ascertain the best-fit PMS value that aligns with the observed structural parameters of the cluster. 
As detailed in Table \ref{tab:initial-conditions-ABC-table2}, two PMS values (S=1 and S=0) were considered for each initial total mass of the star cluster.

By varying the initial stellar mass of the pre-ONC object ($M_{\star}$, column 3 of Table \ref{tab:initial-conditions-ABC-table2})  within the range  $\approx 1000 \msun$ to $\approx 4000 \msun$, we aim to identify the model that best matches the observed present-day values of mass, number of stars, r$_h$, and r$_t$ for the ONC, Pleiades, and Hyades at their reported ages. Since the star formation efficiency (SFE) in all models is $\approx$0.33, the total mass of the embedded cluster can be estimated as $M_{\rm \star+gas} \approx 3 M_{\star}$.

The \textsc{Nbody6} version used here incorporates the tidal effects of the host galaxy by modeling three key components of the Milky Way: the bulge, halo, and disk. These tidal effects are accounted for using an analytical galactic background potential. The bulge is modeled as a point source with a potential of the form:
\begin{eqnarray}
\Psi_{\rm b} \left(r\right) = -\frac{G M_{\rm b}}{r},\label{equ:bulge}
\end{eqnarray}
where M$_{\rm b} = 1.5 \times 10^{10} \msun$ is the mass of the bulge component. The \cite{Miyamoto75} model for the disk is:
\begin{eqnarray}
\Psi_{\rm d} \left(x,y,z\right) = -\frac{G M_{\rm d}}{\sqrt{x^2+y^2+\left(a + \sqrt{z^2+b^2}\right)^2}},\label{equ:disk}
\end{eqnarray}
where $a = 4$ kpc and $b = 0.5$ kpc are the disk scale length and scale height, respectively, and
M$_d = 5 \times 10^{10} \msun$ is the total mass of the disk component. The phantom halo is represented by a logarithmic potential given by:
\begin{eqnarray}
\Psi_{\rm h}(R)= \frac{v_0^2}{2} ~ \ln(R^2+R_c^2),\label{equ:halo}
\end{eqnarray}
where the velocity $V_0=220\kms$ represents the speed of an object in a circular orbit at $ R_\mathrm{c}=8.5\kpc$ from the Galactic center \citep{Allen91, Aarseth03, Madrid12}. 

The parameters for the Milky Way-like potential in all models are based on \cite{Allen91} and \cite{Xue08}. As shown in column 2 of Table \ref{tab: observational-results-OPH-table1}, the three SCs studied are among the closest neighbors of the Solar System, and their distances from the Sun are negligible compared to the Galactocentric distance (d $<<$ 8.5 kpc). Therefore, a Galactocentric distance of 8.5 kpc is assumed in all models (for details, see Table \ref{tab: observational-results-OPH-table1}).

For the initial number of stars in our models, we considered two different values: one with approximately $10^4$ stars plus brown dwarfs, similarly as assumed by \cite{Kroupa00} and \cite{KroupaAH01} for the ONC population, and another with around 4000 stars plus brown dwarfs (model M1.5k-S1).

\color{red}
\begin{table}
\centering
\caption{Details of all simulated star clusters, each starting with different initial parameters. Column 1 lists the model name, which includes 18 initial cluster models. The first three models are based on expanding clusters as described by \citet{Kroupa00} and \citet{KroupaAH01}. Columns 2 and 3 provide the initial number of stars plus brown dwarfs and the stellar plus brown dwarfs mass of the clusters at birth. Column 4 gives the initial half-mass radius ($r_{hi}$). Column 5 shows the degree of initial mass segregation in the star clusters. The final column presents  the cluster model, indicating the selected initial distribution of stars for the setup. Two different models were used: the Plummer model (P) and King's model (K), with the King model's $W_0$ parameter also provided. }
\begin{tabular}{cccccc}
\hline
\begin{turn}{0}Name\end{turn}&$N_{in}$& $M_{\star}$&$r_{hi}$&$S$ &Cluster\\
&$\times 10^3$ & [M$_\odot$] & [pc] &&model\\
\hline
M3.6k-S1&9.8&3642&0.45&1&P\\
M3.7k-S1&9.8&3731&0.2 &1&P\\
M3.7k-S1-K&9.8&3702&0.45&1&K $(W_0=7)$\\
M3.9k-S1&10&3924& 0.3&1&P\\
M3.8k-S0&10&3793& 0.29&0&P\\
M1.5k-S1&4.2&1556&0.26&1&P\\
M1k-S1&2.9&1000&0.245&1&P\\
M1k-S0&2.8&1000&0.245&0&P\\
M1.2k-S1&3.1&1200&0.251&1&P\\
M1.2k-S0&3.7&1200&0.251&0&P\\
M1.4k-S1&3.8&1400&0.256&1&P\\
M1.4k-S0&3.9&1400&0.256&0&P\\
M1.6k-S1&4.4&1600&0.26&1&P\\
M1.6k-S0&4.3&1600&0.26&0&P\\
M1.8k-S1&4.9&1800&0.265&1&P\\
M1.8k-S0&5.2&1800&0.265&0&P\\
M2k-S1&5.9&2000&0.267&1&P\\
M2k-S0&5.7&2000&0.267&0&P\\
\hline
\end{tabular}
\label{tab:initial-conditions-ABC-table2}
\end{table}
\color{black}

\section{Results}\label{Sec:Results}

We now present the results of our simulations, as shown in Table \ref{tab:initial-conditions-ABC-table2}. We investigated how the initial cluster mass, half-mass radius, and initial segregation influence the evolution of the ONC-like star cluster. Then these results are compared with the current observational characteristics of such clusters. The observational constraints include the number of stars, total masses, half-mass radii, and tidal radii of the star clusters, which are detailed in Section \ref{Sec: Observations} and summarized in Table \ref{tab: observational-results-OPH-table1}.  In Figures 1 to 5, the simulated cluster models are depicted by colored lines, each distinguished by a unique line style. The symbols represent observational data: triangles for the ONC, circles for the Pleiades, and squares for the Hyades.  Further details on the observational data can be found in Table \ref{tab: observational-results-OPH-table1}.

\subsection{Evolution of the tidal radius}\label{SubsubSec:evolution-tidal-radii}

\begin{figure*}
\includegraphics[width=15 cm]{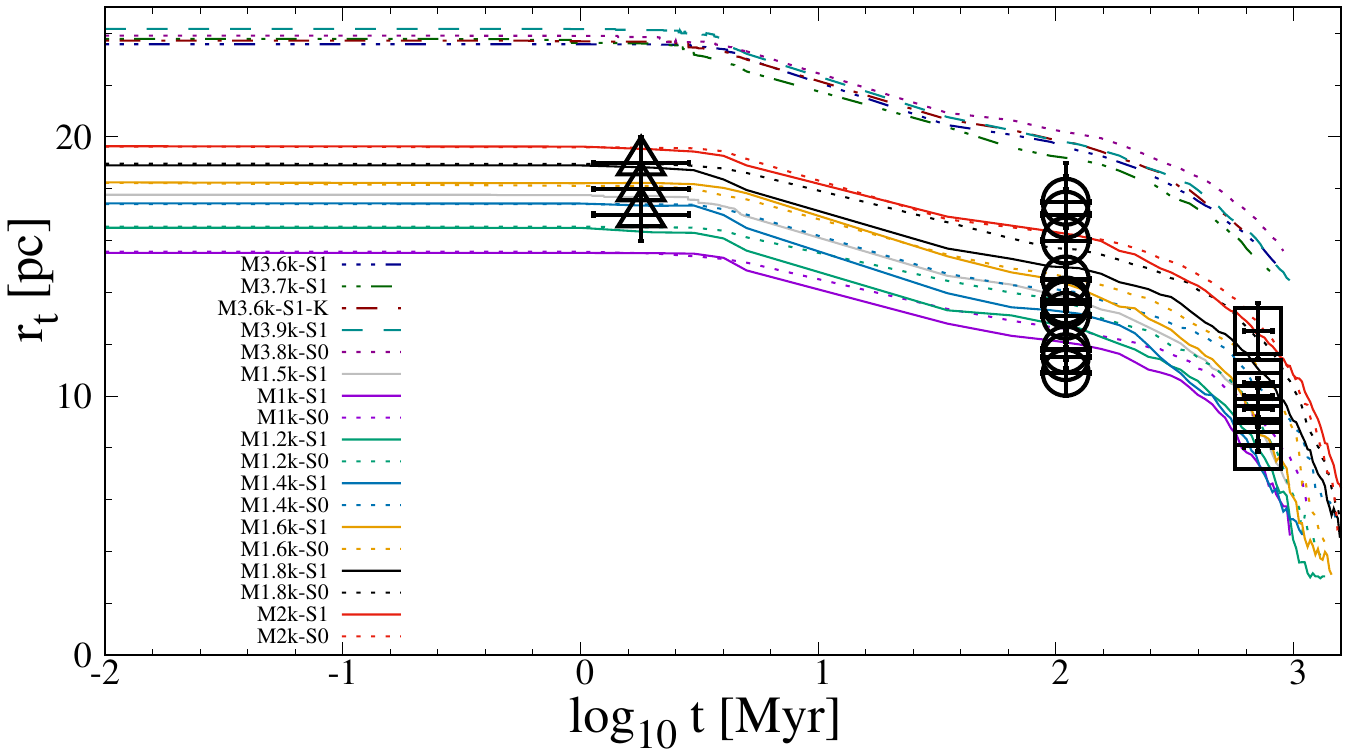}
\caption{Time evolution of the tidal-radii ($r_t$) for different cluster models listed in Table \ref{tab:initial-conditions-ABC-table2}.   The colored lines represent the simulated cluster models, with specific line styles distinguishing each model. }
\label{fig:plot-rt-all-OPH}
\end{figure*}

In our models, only the stars within the theoretical tidal radius are considered for calculating the half-mass, total mass, core radius, and total number of stars. We calculate the tidal radius of the modeled cluster at each time step and compare it with the observed values corresponding to the age of each star cluster.   The theoretical calculation of the star cluster tidal radius, $r_t$, is determined from  
\begin{eqnarray}
r_{\rm t} = R_{\rm G} (\frac{M}{3~M_{\rm G}})^{1/3},
\label{equ:rt}
\end{eqnarray}
where  $R_G =8.5~$kpc is the galactocentric distance of the cluster, $M_{\rm G} = 5 \times 10^{10}~\msun$ is the mass of the galaxy, and  $M$ represents the mass of the star cluster enclosed within the tidal radius.

Fig. \ref{fig:plot-rt-all-OPH} shows the time evolution of the tidal radii for all models. As the mass of the star clusters steadily decreases in all models, this leads to a corresponding reduction in the tidal radii.  Fig. \ref{fig:plot-rt-all-OPH} demonstrates that initial full mass segregation (S = 1, solid lines) correlates with a higher rate of mass loss across all models. This, in turn, drives a faster reduction in tidal radius ($r_t \propto M_c$), as evidenced by the comparison with initially unsegregated models (S = 0, dotted lines).
The open triangles, circles, and squares represent the observed tidal radii for the ONC, Pleiades, and Hyades, respectively, as reported in various literature sources; see Table \ref{tab: observational-results-OPH-table1}.

\subsection{The evolution of the total number of stars and stellar mass}\label{SubSec:evolution-n-m}

\begin{figure*}
\includegraphics[width=15 cm]{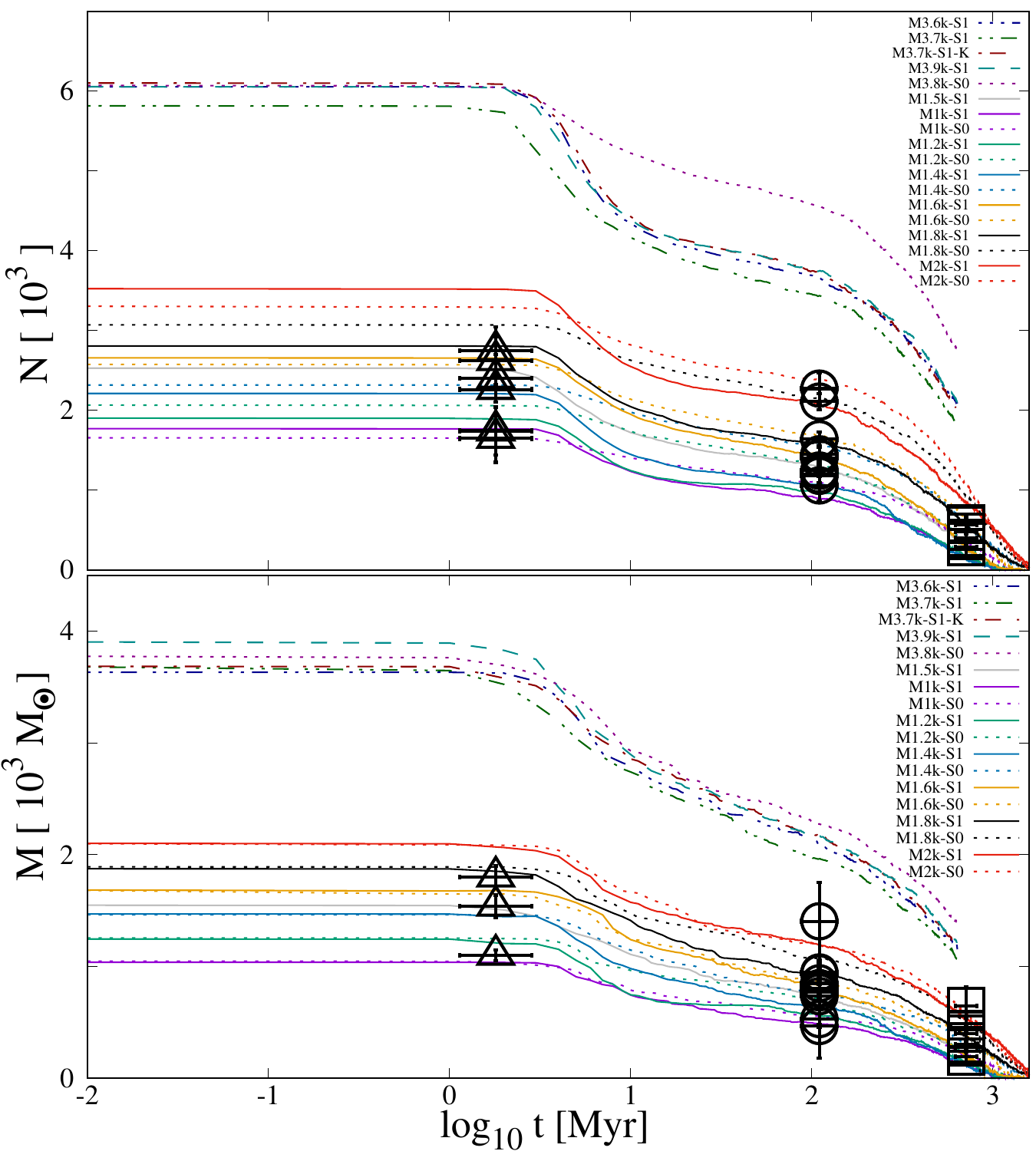}
\caption{The time-evolution of the total number of stars (top panel) and total mass (bottom panel) within $r_{\rm t}$for different cluster models.}
\label{fig:plot-n-all-OPH}
\end{figure*}

Figure \ref{fig:plot-n-all-OPH} (top panel) shows the number of stars within the tidal radius ($r_t$) for each model. The open triangles, circles, and squares represent various observed values for the number of stars in the ONC, Pleiades, and Hyades, respectively, taken from the literature; see Table \ref{tab: observational-results-OPH-table1}.

Observational limitations, including magnitude thresholds and instrument sensitivity, hinder the detection of faint objects such as brown dwarfs. To ensure a valid comparison, we restricted our simulation analysis to stars with masses greater than 0.1 $M_\odot$ for counting purposes.   As shown, none of the first five models listed in Table \ref{tab:initial-conditions-ABC-table2}, starting with about $10^4$ stars, which include about 6000 stars more massive than 0.1 $M_\odot$, reduced the stellar population enough to match the present-day observed values. Models without PMS exhibited both a slower decrease in the number of stars, indicating longer survival times, compared to the other models (Fig. \ref{fig:plot-n-all-OPH}).

To resolve this discrepancy, model M1.5k-S1 is designed to match both the observed number of stars and the luminous mass of the ONC. This model assumes an initial star count of $N=4200$, corresponding to a stellar plus brown dwarfs mass of $M_{*} \approx 1500\msun$, and a residual gas mass of $M_{gas} \approx 3000\msun$ to account for the GE effect. It shows the most favorable evolutionary trend among our models for reproducing the stellar populations of the ONC, the Pleiades, and the Hyades. Additionally, the initial number of stars ($m> 0.1 M_{\odot}$) in this model is consistent with previous studies that simulated the dynamical evolution of the ONC and its runaway stars using \textsc{Nbody6} \citep{Scally05, Subr12,Pavlik19}.

Star clusters lose mass over time due to four main factors: the GE effect, the stellar evolution of individual stars, evaporation driven by the two-body relaxation process, and the tidal field, which arises from interactions with the parent galaxy. These processes drive the transition from dense, embedded clusters via an open cluster phase to their eventual dissolution. In Figure \ref{fig:plot-n-all-OPH} (bottom panel), the total mass of the star clusters within $r_t$ is plotted as a function of time. The open triangles, circles, and squares represent the observed values for the total stellar mass in the ONC, Pleiades, and Hyades (Table \ref{tab: observational-results-OPH-table1}).  

Due to the weakening of the gravitational potential well caused by early GE and rapid mass loss from the evolution of massive stars, all models undergo violent expansion and lose a portion of their stars. This initial ejection of gas leads to a positive mean radial velocity of the stars, causing many of them to move outward. Stars with high enough energies become unbound from the cluster, determining the amount of mass lost. Less energetic stars move outward while staying bound.  Therefore, the evolution of mass in our models within the first few Myr is dominated more by the GE effect and by stellar evolution of heavy-mass stars.

\subsection{Evolution of the half-mass radius}\label{SubsubSec:evolution-rh}

Using the models described above, we now discuss the evolution of the half-mass radii of stars bound to the clusters,  starting with different initial conditions and undergoing early GE (Figure \ref{fig:plot-rh-all-OPH}).  The open triangles, circles, and squares represent the observed $r_h$ values of the ONC, Pleiades, and Hyades, respectively (see Table \ref{tab: observational-results-OPH-table1}). All models expand as a result of the dilution of the gravitational potential well, which is driven by the early GE and rapid mass loss due to the evolution of the massive stars. 
Initially, $r_h$ remains constant until $t_D = 0.6$ Myr (ultracompact $H_{\rm II}$ phase). Afterward, the gas mass evolves following Equation (\ref{equ:gas-mass}).  Significant mass loss occurs around $t = 6$ Myr, corresponding to the lifetime of a 50 $M_\odot$ star, which is the maximum stellar mass in our models. This lifetime is derived using the SSE package used in this study (see Section \ref{Sec:Description of the models}). At this point, stellar evolution and Type II supernova explosions mark the onset of the primary mass-loss phase.  Following this period, two-body relaxation and the tidal effects of the parent galaxy drive a more gradual decrease in cluster mass for the remainder of its lifetime.

The evolution of the half-mass radius shows good agreement with the observational constraints for almost all models. However, models based on the Marks-Kroupa relation show slightly better agreement with the observational data compared to models M3.6k-S1, M3.7k-S1, and M3.7k-S1-K. Consistent with the expanding cluster models proposed by \citet{Kroupa00} and validated by \citet{KroupaAH01, Kroupa13}, the half-mass radius ($r_h$) increases due to mass loss in all cases, aligning well with observations (Fig. \ref{fig:plot-rh-all-OPH}). As illustrated in the top panel of Fig. \ref{fig:plot-rh-all-OPH}, models initialized with PMS (S = 1) display larger $r_h$ values compared to those without (S = 0).
Model M3.8k-S0, which starts without PMS, shows a slightly lower $r_h$ compared to the median observational values, particularly in the case of the Hyades. As illustrated in Fig. \ref{fig:plot-rh-all-OPH}, models M3.6k-S1, M3.7k-S1, M3.7k-S1-K, M3.9k-S1 and M3.8k-S0, with higher initial masses, survive for more than 1 Gyr, while model M1.5k-S1 dissolves at around 1 Gyr. In the case of model M1.5k-S1, the age of the Hyades is near its dissolution time, and its half-mass radius decreases slightly as it approaches dissolution.

 \subsection{Evolution of the core radius}\label{sub:result-rc-rvolution} 

The core radius ($r_{\rm c}$) of a star cluster is the region with the highest stellar density, offering insights into the dynamical state and evolution of the cluster. Its evolution is driven by dynamical interactions, mass loss, and the presence of stellar remnants such as black holes. Processes such as two-body relaxation, stellar evolution, and mass segregation drive changes in the core radius.

Observationally, the core radius is defined as the distance at which the surface density of stars in a cluster drops to half of its central value ($\Sigma(r_{\rm c}) = 0.5~\Sigma(0)$). To calculate the core radius, a fitting model is used, with King’s definition being a common example \citep{King62}. This method involves two main steps: measuring the surface number density and identifying the radius where it falls to half of the central value. The empirical surface density function of \cite{King62} is frequently used to model the stellar distribution, as it closely resembles the observed surface density ($\Sigma_s$) distributions in star clusters. King's function is defined as:
\begin{equation}\label{equ:sigma-king} \Sigma_s(r) = k' \left( \frac{1}{\sqrt{1+y}} - \frac{1}{\sqrt{1+y_t}} \right)^2, \end{equation}
where $y = \left(\frac{r}{r_c}\right)^2$, $y_t = \left(\frac{r_t}{r_c}\right)^2$, $k'$ is a normalization constant related to the central density, and $r_t$ is the tidal radius of the cluster.

\begin{figure*}
\includegraphics[width=14 cm]{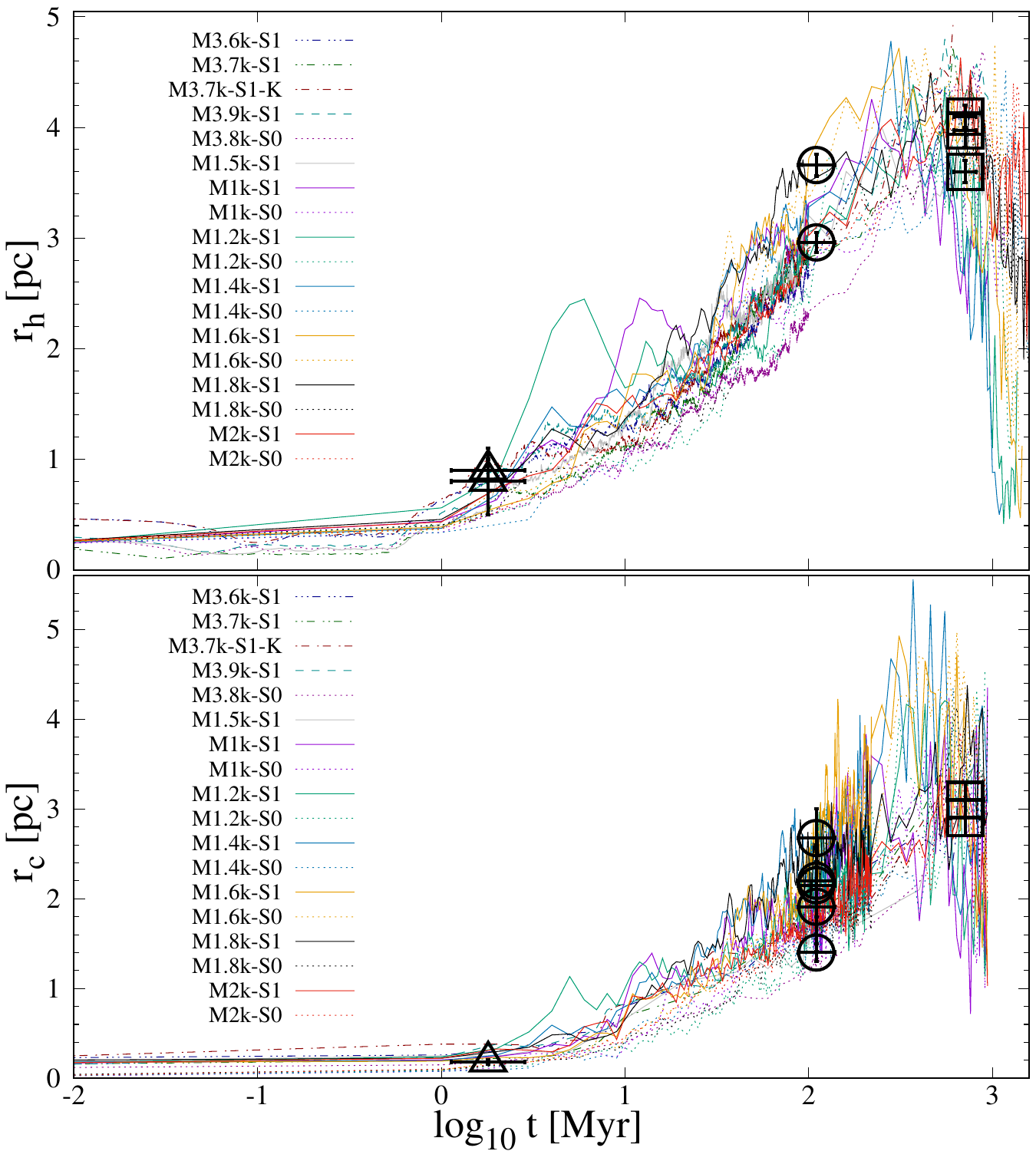}
\caption{The evolution of 3D half-mass radii ($r_h$, top panel) and the core radii ($r_c$, bottom panel) of simulated star clusters over time. }
\label{fig:plot-rh-all-OPH}
\end{figure*}

The results of our simulations are presented as the line types calculated using the observational method, as shown in Fig. \ref{fig:plot-rh-all-OPH} (bottom panel). The open triangles, circles, and squares represent the core radius of the ONC, Pleiades, and Hyades, respectively, as reported in various studies; see Table \ref{tab: observational-results-OPH-table1}. As a star cluster evolves, the core radius generally expands due to a combination of mass loss from stellar winds and supernovae, as well as gravitational interactions between stars, particularly in the dense central regions. Two-body relaxation redistributes energy, causing core contraction as some stars gain energy and move outward while others lose energy and sink toward the center. Initially, mass segregation, where more massive stars migrate toward the core, also drives core contraction. However, subsequent dynamical interactions between massive and less massive stars may lead to core expansion. Stellar evolution further influences the core radius, as mass loss from aging stars, such as through supernovae or stellar winds, reduces the cluster's gravitational binding energy, which can lead to core expansion.  As can be seen in Fig. \ref{fig:plot-rh-all-OPH} (bottom panel), it is evident that the core radius of the modeled clusters along their evolutionary tracks closely matches the observational values for the three clusters.

It is worth mentioning that the empirical  mass-radius relation 
(Eq. \ref{equ:marks-kroupa-relation}) proposed by \citet{Marks12}, suggests only a weak dependence of the initial half-mass radius on the initial mass of the cluster. In our simulations, the variation in initial cluster mass ranging from approximately 1000 to 4000 $M_{\odot}$ is relatively modest (Fig. \ref{fig:plot-rh-all-OPH}). As a result, the expected differences in the initial half-mass radii are small, which leads to a similar dynamical evolution of the core and half-mass radii across the different models.   Given the small range of initial masses and half-mass radii (0.2-0.3 pc) in our models, their direct impact on the subsequent evolution of $r_h$ and $r_c$ is expected to be limited.  In contrast, the tidal radius is more directly and strongly influenced by the total cluster mass, given its dependence on the external Galactic potential. Therefore, differences in mass are clearly reflected in the evolution of the tidal radius, consistent with what is observed in our models (Fig. \ref{fig:plot-rt-all-OPH}).

\section{Conclusions}\label{Sec:Conclusions}

In this study, we successfully reproduced the dynamical evolution of the ONC, Pleiades, and Hyades, suggesting that the Pleiades may have been similar to the ONC about 100 Myr ago and will be similar to the Hyades around 700 Myr in the future. 

This work establishes a non-causal connection between the ONC and the Pleiades/Hyades, suggesting that Galactic clusters initially form with large core radii, filling their tidal radii after gas loss from a compact state similar to the ONC. They likely also form as expanding OB association nuclei.

Through $N$-body simulations using the NBODY6 code, we explore the evolutionary pathways of the ONC as the progenitor of the Pleiades and Hyades. Our simulations show strong agreement with the observed cluster properties, such as total mass, stellar population, and half-mass, core, and tidal radii (Table \ref{tab: observational-results-OPH-table1}). Models initialized with a canonical IMF and sizes from the Marks-Kroupa radius-mass relation particularly match observational data for the ONC, Pleiades, and Hyades, with Model "M1.5k-S1" showing the best fit. 

Models closely matching observational data are produced by initial stellar populations of approximately $3.8~ \times~10^ 3$ to $5.2~\times~10^3$ stars plus brown dwarfs (with total initial masses ranging from 1400 to 1800 $\msun$). Our simulations reveal substantial mass loss over time, with clusters losing 50-60 per cent of their mass in 110 Myr and 70-85 per cent in 794 Myr, consistent with the observed evolution of the Hyades.  We did not explicitly study mass segregation in the models and resort to the $N$-body modeling by \citet{Pavlik19} for the ONC. At gaes of 100 Myr and older, it is already well established clusters are dynamically mass segregated.

In conclusion, our simulations demonstrate that $N$-body models with well-constrained initial conditions can accurately reproduce the dynamical evolution of star clusters at different evolutionary stages. This work highlights the importance of combining simulations and observational data to refine our understanding of star cluster formation and evolution.  The rather astounding result that all three clusters are conspicuously located in a very similar region of the sky may arise from the transition in the mass function of embedded clusters from a power-law distribution to that of post-gas-expulsion clusters that has a peak near 1000 $\msun$ \citep{KroupaBoily2002}.

\section*{Acknowledgements}
The authors dedicate this paper to Sverre Johannes Aarseth, (20 July 1934 – 28 December 2024), whose profound contributions to star cluster research, particularly through the development of his long-lived N-body code, have shaped and advanced our understanding of stellar dynamics.  AHZ acknowledges support from the Alexander von Humboldt Foundation. HH is grateful to the staff at the Helmholtz-Institut für Strahlen- und Kernphysik (HISKP) and Argelande Institute for Astronomy (AIfA) for their hospitality and acknowledges financial support from the SPYDOR group at the University of Bonn. PK thanks the DAAD Bonn-Prague Eastern European Exchange Program. 
\section*{Data availability}
The data underlying this article are available in the article.
\bibliographystyle{mnras}
\bibliography{MyReferences} 
\bsp	
\label{lastpage}
\end{document}